\documentclass[lettersize,journal]{IEEEtran}
\pdfoutput=1
\usepackage{hyperref}
\usepackage{setspace}
\usepackage{amsmath,amsfonts}
\usepackage{algorithmic}
\usepackage{algorithm}
\usepackage{array}
\usepackage{CJK}
\usepackage[caption=false,font=normalsize,labelfont=sf,textfont=sf]{subfig}
\usepackage{textcomp}
\usepackage{stfloats}
\usepackage{url}
\usepackage{verbatim}
\usepackage{graphicx}
\usepackage{cite}
\usepackage[justification=centering]{caption}
\begin{document}

\title{Distributed Collaborative Inference System in Next-Generation Networks and Communication}

\author{
    Chuan~Zhang,~\IEEEmembership{Member,~IEEE,}
    Xixi Zheng, 
    Xiaolong Tao,
    Chenfei Hu,
    Weiting Zhang,~\IEEEmembership{Member,~IEEE,} \\and Liehuang~Zhu,~\IEEEmembership{Senior~Member,~IEEE}

\thanks{Chuan Zhang, Xixi Zheng, Xiaolong Tao, Chenfei Hu, and Liehuang Zhu are with the School of Cyberspace Science and Technology, Beijing Institute of Technology, Beijing 100081, China. E-mail: \{chuanz, bit-zhengxixi, 3120231250, chenfeih, liehuangz\}@bit.edu.cn.}

\thanks{Weiting Zhang is with the School of Electronic and Information Engineering, Beijing Jiaotong University, Beijing 100044, China. E-mail: wtzhang@bjtu.edu.cn.}

\thanks{Weiting Zhang is the corresponding author.}

\thanks{This work was financially supported by the National Natural Science Foundation of China (Grant No. 62472032), the Open Project Funding of Key Laboratory of Mobile Application Innovation and Governance Technology, Ministry of Industry and Information Technology, under Grant 2023IFS080601-K, the Young Elite Scientists Sponsorship Program by CAST (Grant No. 2023QNRC001), Xiaomi Research Fund for Young Scholars, and the Fundamental Research Funds for the Central Universities under Grant 2024CX06034.}


} 

\maketitle

\begin{abstract}
With the rapid advancement of artificial intelligence, generative artificial intelligence (GAI) has taken a leading role in transforming data processing methods. However, the high computational demands of GAI present challenges for devices with limited resources. As we move towards the sixth generation of mobile networks (6G), the higher data rates and improved energy efficiency of 6G create a need for more efficient data processing in GAI. Traditional GAI, however, shows its limitations in meeting these demands. To address these challenges, we introduce a multi-level collaborative inference system designed for next-generation networks and communication. Our proposed system features a deployment strategy that assigns models of varying sizes to devices at different network layers. Then, we design a task offloading strategy to optimise both efficiency and latency. Furthermore, a modified early exit mechanism is implemented to enhance the inference process for single models. Experimental results demonstrate that our system effectively reduces inference latency while maintaining high-quality output. Specifically, compared to existing work, our system can reduce inference time by up to 17\% without sacrificing the inference accuracy. 

\end{abstract}

\begin{IEEEkeywords}
Generative artificial intelligence, early exit, collaborative inference, next-generation networks and communication.
\end{IEEEkeywords}

\section{Introduction}

\begin{figure}[tp]
    \centering
    \includegraphics[width=8.5cm]{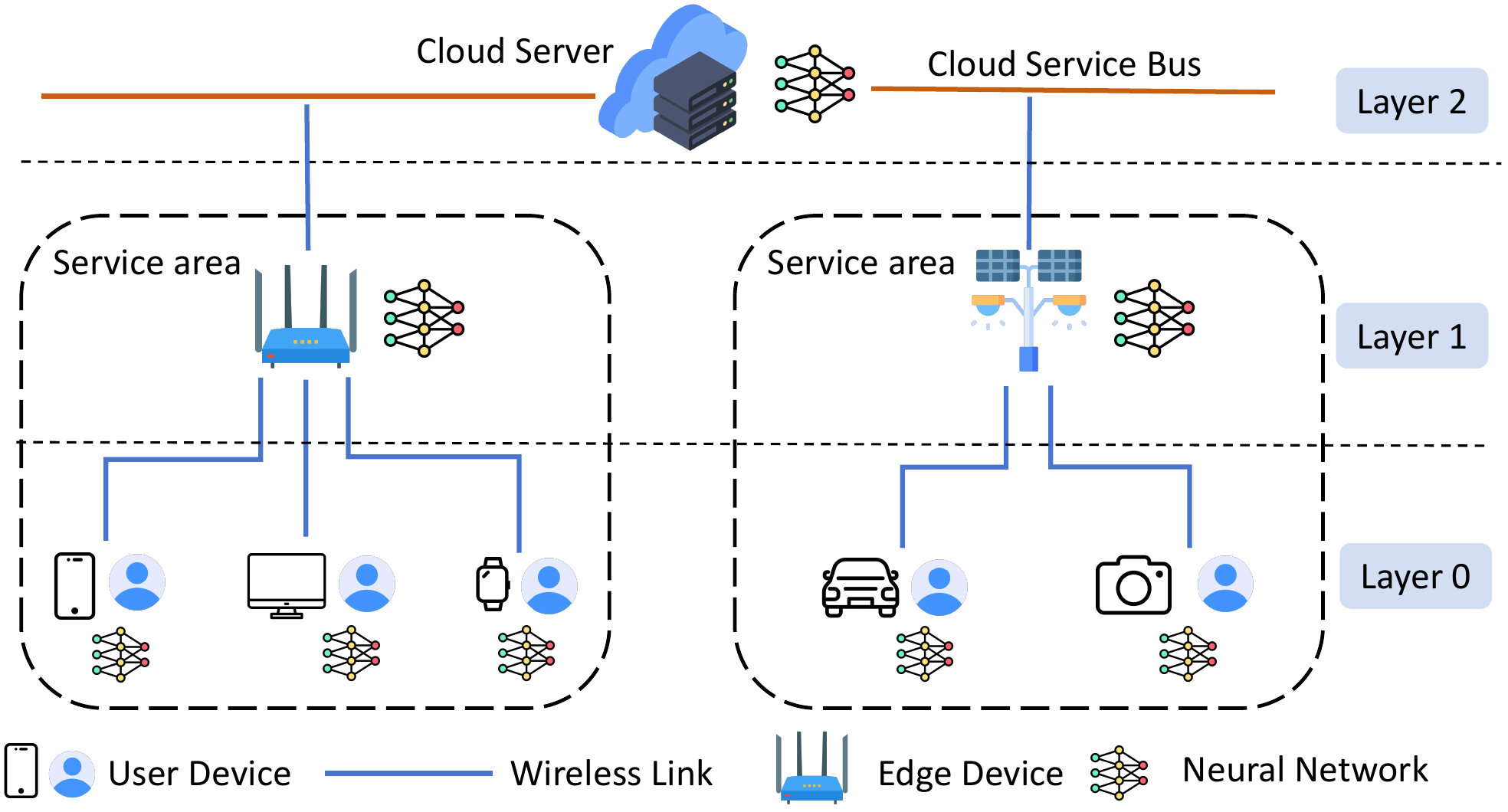}
    \caption{Examples of services provided by multi-level collaborative inference in mobile networks.}
    \label{fig:1}
\end{figure}

\IEEEPARstart{G}{enerative} artificial intelligence has revolutionized the field of artificial intelligence, showing exceptional potential in applications such as sentiment analysis, machine translation, and text generation. Models based on the Transformer architecture, such as the GPT series, have significantly propelled GAI technology forward. These models have not only improved inference performance but have also expanded the scope and depth of GAI applications. For example, applications of discriminative task like sentiment analysis\cite{xu2022systematic},\cite{prasad2022machine}, fake news identification\cite{athira2023systematic},\cite{khanam2021fake}, user feedback analysis\cite{xu2017new}, and natural language inference\cite{long20246g} have demonstrated superior performance with the Transformer architecture. With the progressive integration of 6G mobile networks into society, the integration of discriminative tasks with GAI becomes increasingly critical. However, the enhanced data transmission speeds, reduced latency, high reliability, and extensive connectivity proposed by 6G \cite{letaief2019roadmap},\cite{yang2020artificial} pose challenges for the efficient execution of discriminative tasks.

When deploying GAI models in practice, the challenges primarily stem from the substantial computational resource requirements and communication latency. The need for high computational resource makes it difficult to implement these models on resource-constrained devices such as smartphones or embedded systems\cite{li2024flexnn},\cite{capotondi2020cmix},\cite{li2021model}, and also limits their feasibility in real-world scenarios\cite{desislavov2023trends},\cite{schwartz2020green},\cite{georgiou2022green}. Efficient deployment is crucial to meet the demands of low-latency responses. Consequently, some users resort to cloud computing services to handle complex inference tasks\cite{banitalebi2021auto},\cite{yao2022edge},\cite{zhao2024socialized}. While cloud platforms provide the necessary computational resources and infrastructure to support the deployment of large-scale AI models, this approach introduces new problem, particularly related to communication latency. This kind of latency, along with model computational latency, collectively constitutes inference latency. The delay caused by data transmission between the user's device and the cloud server as well as model computation can sometimes be unacceptable, especially for time-sensitive applications\cite{naha2020deadline},\cite{jayaraman2017analytics},\cite{qiao2024edgeoptimizer}. 

In order to solve the above problems, many schemes have been proposed in recent years. Cocktail\cite{gunasekaran2022cocktail} used integrated learning of multiple small models to reduce computational latency through parallel execution. It also dynamically adjusted model integration to minimise computational cost. Although this work achieved low computational latency and cost through collaborative inference with small models, it could not achieve fine-grained model selection, which may lead to accuracy loss. To address this issue, Wang et al. \cite{wang2023tabi} proposed a system called Tabi with a multi-level inference engine that used small models and optional Large Language Models (LLMs) to provide inference capabilities for demanding tasks. However, these works did not consider the impact of communication latency on inference, meaning they did not take into account inference scenarios in mobile networks and communication. As for the accelerated inference of the model itself, Zhou et al. \cite{zhou2020bert} proposed a new Patience-Based Early Exit (PABEE) mechanism that allows the model to dynamically stop inference. Inspired by biological vision mechanisms, Jiang et al. \cite{jiang2020learning} designed a layer-skippable network that can dynamically classify objects from coarse to fine. Different from previous works, to make the layer skipping strategy independent of input samples, Liu et al. \cite{liu2024accelerating} proposed a unified layer-skipping strategy, where the number of layers to be skipped from the computation is selected based only on the target speedup ratio, and then the corresponding number of intermediate layer computations are skipped in a balanced manner. Although these studies have explored early exit mechanisms, these acceleration schemes lack specificity, meaning they are not tailored for accelerating certain inference tasks in next-generation networks and communications. Therefore, overall, deploying efficient inference systems in mobile network and communications remains challenging.

Given this situation, we propose a multi-level inference system that makes use of key algorithms such as attention-based pruning. Our approach features a ``cloud-edge-end" multi-level inference framework, which integrates a ``early exit" in order to ensure efficient task completion with minimal inference latency (\autoref{fig:1}). We design a decision algorithm to guide the offload settings of the end device. In addition, we optimizes the early exit mechanism in the transformer architecture to enhance its adaptability to inference tasks. This system aims to balance the trade-offs between computational demands and inference quality, optimising the deployment strategy for diverse network layers. The main contributions of this paper are as follows:

\begin{itemize}
\item[$\bullet$]
We have designed a distributed multi-level inference framework suitable for next-generation networks and communication. This framework is capable of ensuring inference quality while maintaining low overall latency.
\item[$\bullet$]
Our system features a task offloading strategy designed to optimise the efficiency and latency of the deployment model system. Specifically, our task offloading strategy employs a special confidence algorithm that can make offloading decisions more rational, and attention-based pruning that reduces communication and re-inference delays.
\item[$\bullet$]
We implement an enhanced early exit mechanism tailored for single-model inference, improving overall system performance and reducing inference latency. Concretely, we design an early exit algorithm for discriminative tasks adapted to the transformer model.
\item[$\bullet$]
We build a prototype of the framework using the Python language and demonstrated the system's effectiveness through experiments. Compared to existing work, our system can reduce inference time by up to 17\% without sacrificing the inference accuracy.
\end{itemize}

\section{SYSTEM MODEL AND PROBLEM FORMULATION}
In this paper, we focus on service offloading in hierarchical mobile networks. In this scenario, the network is ideally divided into a three-level structure (in practice, it may be divided into fewer or more levels, with the three levels not losing their universality). Each level has a corresponding physical device with a model on it that matches its computational resources. As the level increases, the computational resources of the physical devices become more abundant, and the computational latency and accuracy of their models also increase relatively. In this section, we present the deployment logic of the model (Part \textit{A}), followed by a description of the system's workflow (Part \textit{B}) as well as a modular presentation of the system model (Parts \textit{C}, \textit{D}, and \textit{E}), and finally the problem formulation (Part \textit{F}).

\subsection{Deployment Logic}
Our system leverages a distributed deployment of multiple models across different levels of the mobile network, including the cloud, edge, and end devices. In real-world applications, the heterogeneous distribution of computational resources means that not all devices in the network can support large, computationally intensive GAI models. To address this challenge, it is crucial to match the models deployed at each network level with the specific computing capabilities of the corresponding devices. This requires a thorough profiling of each model to assess key factors such as computational latency, accuracy, and resource consumption.

At higher levels of the network, such as cloud servers with abundant computational resources, larger and more complex models are deployed, as they can leverage the higher processing power and storage capacity. Conversely, at lower levels, such as user devices with limited resources, smaller models are utilized to ensure responsiveness and efficiency without overburdening the devices.

The deployment decision is driven by several considerations, including the trade-offs between model accuracy, computational demands, and latency. While larger models generally provide higher accuracy, they also require significantly more computational power and can introduce higher latency, which may not be acceptable in time-sensitive applications. Therefore, the placement of models is guided by a careful balancing of these factors, optimizing for both performance and resource utilization to ensure the system operates efficiently across varying network conditions.


\subsection{Inference Workflow}
\begin{figure}[tp]
    \centering
    \includegraphics[width=9cm]{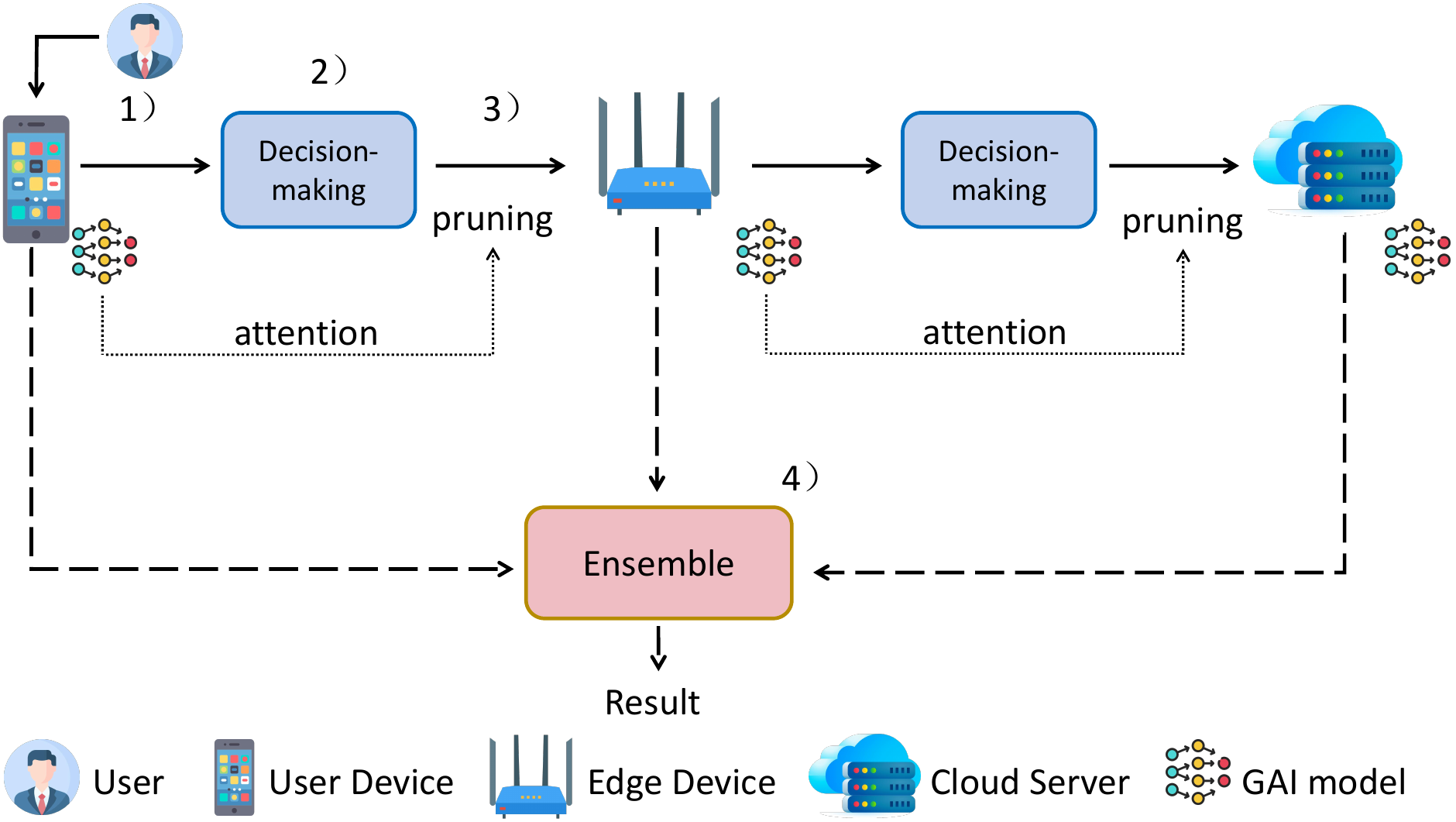}
    \caption{Inference workflow of our system.}
    \label{fig:2}
\end{figure}

As shown in \autoref{fig:2}, the workflow of a three-level system consists of the following  steps:
\begin{itemize}
\item[1)]
The user presents the inference task to the user-side model and gets the result of the initial inference.
\item[2)]
A decision is made regarding the necessity to offload the inference task to a higher-level model based on the initial inference results and intermediate values.
\item[3)]
If offloading is required, the task request is pruned based on attention mechanisms and then input to the next level model. This inference and offloading decision process is repeated in next level.
\item[4)]
The inference process ends when a decision no longer needs to offload tasks to the higher-level model, or when the final-level model is reached. The aggregation of the results of each inference in the process according to weights (determined by the accuracy of the model itself) is the final result. The results are then returned to the user.
\end{itemize}

This workflow performs inference tasks efficiently and is able to leverage the strengths of each level of the model, balancing workloads between local devices and cloud servers while optimising time, resource and accuracy costs.

\subsection{Task Offloading Dispatcher}
The task offloading dispatcher in our system consists of two main modules, the confidence-based probabilistic offloading module and the attention-based pruning module.

\textbf{Confidence-based probabilistic offloading.} GAI should be able to indicate the correctness of inference when dealing with complex natural language applications.In our system, returning inference results from small models too early can lead to a faster inference process. However, incorrect results risk violating the accuracy goal and may negatively impact the overall accuracy of the collaborative inference. Therefore, our algorithm calculates the confidence level of the model based on the result of a particular inference, which can be interpreted as the model's ``confidence level" in its output. Utilizing this, we introduce probabilistic properties, and for each inference result, we calculate the probability of offloading the task to a higher-level model, which forms the basis of our dispatcher.

\textbf{Attention-based pruning.} After dispatching, more challenging tasks are offloaded to a higher-level model due to the low confidence of the smaller model. This may result in the same task being inferred multiple times, causing tail latency. Conversely, we find that smaller models can reduce latency. We can leverage the small model's understanding of the task statement to expedite the inference process of the higher-level model by pruning unnecessary words from the input data. This approach helps offset the excess latency overhead caused by multiple inference. To some extent, it also reduces the communication overhead. Specifically, we record the attention values at each level of the inference process and sum them to gauge the model's understanding of the task statement. Each word in the task statement is assigned a weight, and unimportant words are pruned based on a set threshold. The pruned result then becomes the input for the higher-level model.

\subsection{Single-model Inference Acceleration}
For a model at a high level of the network, its inference input is a task statement based on attention pruning of a small model. Thus, it appears that its inference speedup is obtained simply by reducing the length of the input.

\begin{figure}[tp]
    \centering
    \includegraphics[width=9cm]{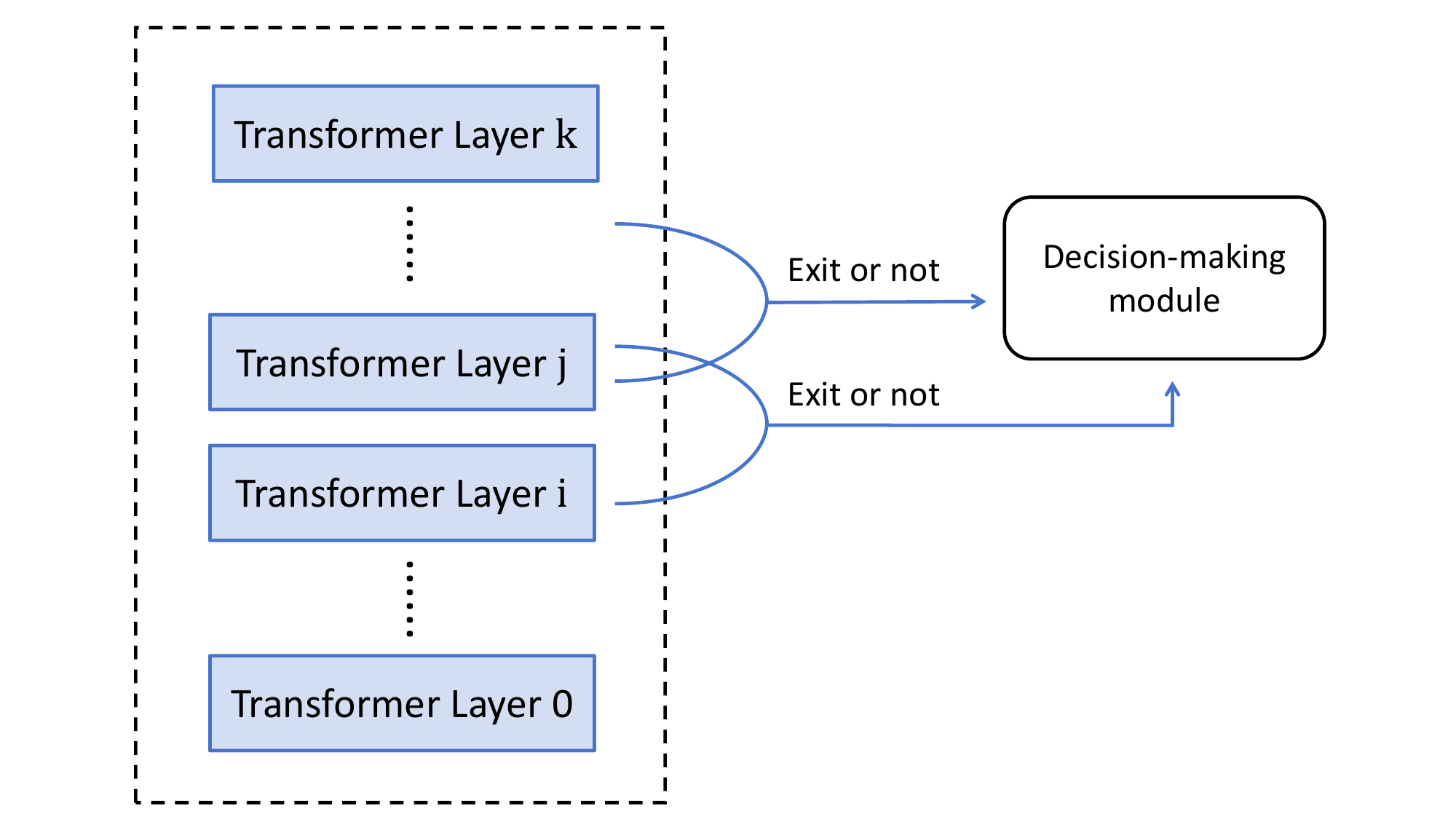}
    \caption{Accelerating of single model inference.}
    \label{fig:3}
\end{figure}

For larger GAIs, models contain many layers and millions or even billions of parameters, making them computationally expensive and inefficient in terms of memory consumption and computational latency. And overparameterisation by stacking large numbers of transformer layers can lead to a problem called ``overthinking" in the inference process. That is, for many inference tasks, their shallow representation in early layers may be sufficient for correct classification without the need for computation in later layers. This overthinking problem leads to a waste of computational resources and is a problem to be addressed in our system. To address this latency, which is mainly caused by the number of model layers, we refer to the work on implementing early exit in models of the transformer architecture, and design acceleration for individual models, i.e. optimised early exit (\autoref{fig:3}). We present each layer of a model's inference with its results in some form, and after the exit condition is satisfied, we treat the results of the exiting layer directly as the final result for that model.

\subsection{Generation of Inference Results}
For the tasks that are offloaded, instead of treating the results of the models at the last level as the results of the whole system, we use weighting to aggregate them with the results of the models at the previous levels to improve the accuracy of the inference results (the weights of the model are determined by the properties of the model itself during pre-profiling). This operation does not incur any additional model inference overhead, as the softmax output is easily accessible during model inference. Existing work has shown that inference of some independent models, even weak ones, in conjunction with some large models can yield more accurate results than the large models alone\cite{krogh1994neural},\cite{maclin1997empirical}.

\subsection{Problem Formulation}
Based on the construction of our system, the user is given an inference task with a target $g(x)$ in addition to specific inference content $x$.


The computational latency $L^{com}$ of an inference task $x$ is defined as the sum of the computational latencies incurred by each individual model in the inference process. Specifically, for the $i$-th model handling task $x_i$, the computational latency $L^{com}(x)$ is defined as follows:

\begin{equation}
L^{com}(x)=\sum_{i=1}^nLat_i^{com}(x_i)
\end{equation}
note that when the first inference is made (i.e., for $ i = 1$, $x_i = x_1 = x$), the task does not need to be pruned.


Since our system is designed for mobile networks, communication latency $L^{tra}$ arises due to the information transfer between different models. The communication latency for transmitting task $x_{j+1}$ between models is denoted as $Lat_j^{tra}(x_{j+1})$ , and it depends on the size of the task and the network transmission rate between layers. Specifically, the total communication latency $L^{tra}(x)$ is given by the following equation:
\begin{equation}
\begin{aligned}
L^{tra}(x)=\sum_{j=1}^nLat_j^{tra}(x_{j+1})\\
=\sum_{j=1}^n\frac{size(x_{j+1})}{S_j}
\end{aligned}
\end{equation}
where $S_j$ denotes the network transmission rate between the $j$-th and the $j+1$-th layers and $size(x_{j+1})$ denotes the size of the task $x_{j+1}$. 

Thus, we can obtain that the total system inference latency for the inference task $x$ is:
\begin{equation}
Latency(x)=L^{com}(x)+L^{tra}(x)
\end{equation}

To maximize the overall inference gain, for inference task $x$, our system goal is formally stated as:
\begin{equation}
\begin{split}
&\min \,\, Latency(x)\\
&s.t.\quad  g(x)
\end{split}
\end{equation}
where $g(x)$ denotes the user's target. It can be understood as a requirement for correctness of inference.

\section{SYSTEM DESIGN}
Based on our system model, this section introduces the offloading based on confidence (Part \textit{A}), the algorithms of attention-based pruning (Part \textit{B}), the ensemble of result (Part \textit{C}), and the optimized early exit (Part \textit{D}).

\subsection{Confidence and Offloading}
A simple way to obtain confidence is to use the softmax probability. In many classification neural networks, softmax is often used as the last layer, taking input values from the previous layer of the network and then converting them into probabilities. The softmax is calculated as follows:
\begin{equation}
softmax(logits_i)=\frac{e^{logits_i}}{\sum_{j=1}^Ke^{logits_j}}
\end{equation}
where $logits$ is the output of the previous layer of the network. The result of the calculation is the probability that class $i$ is in the full set $K$ of classes.

However, some work has shown that many language models suffer from the problem of ``over-confident" in their inference results. Specifically, a model with low accuracy may be highly confident in every inference it makes. Or it may be confident even when it produces incorrect inference results\cite{desai2020calibration},\cite{guo2017calibration}. In other words, if the softmax value is used directly as a confidence indicator in the practical use of model inference, it is likely to carry a risk of loss of accuracy.

To enable our system to make the right decisions about offloading tasks, we introduce temperature scaling in softmax\cite{guo2017calibration}. For the classification task $x$, the confidence is calculated as follows:
\begin{equation}
conf(x)=\mathop{\max}_{i\in K} \frac{e^{logits_i/T}}{\sum_{j=1}^Ke^{logits_j/T}}
\end{equation}
where $T$ is the temperature parameter. Temperature scaling is a very simple calibration method that improves the match between the confidence level and the inference accuracy of the model itself by simply introducing an additional temperature parameter and optimising it on the validation set.

Having obtained the reliable confidence, we can set the threshold. In this way, there is a logic underlying task offloading: if the confidence is below the threshold, the task needs to be offloaded to the high-level network; if the confidence is above the threshold, it does not. However, this cut-off decision can lead to task offloading that is not smooth enough, and does not offload the task when a high confidence result occurs in the small model, which does not take advantage of collaborative inference. So we set up a probabilistic offloading approach when the confidence level is above a threshold, so that high-confidence tasks still have a chance to go to the high-level network for inference. For task $x$, its offloading probability is:
\begin{equation}
p(x)=\left\{
	\begin{aligned}
	\frac{1}{1+e^{k \cdot norm((conf(x)))}} \quad conf(x) > t\\
	1 \quad conf(x) \leq t\\
	\end{aligned}
	\right
	.
\end{equation}
\begin{equation}
norm((conf(x)))=\frac{conf(x)-t}{1-t}-\frac{1}{2}
\end{equation}
where $t$ is the threshold and $k$ is the scaling parameter. Both hyperparameters can be adjusted according to the specific task, which in turn affects the overall correctness of the system.

\begin{figure}[tp]
    \centering
    \includegraphics[width=7cm]{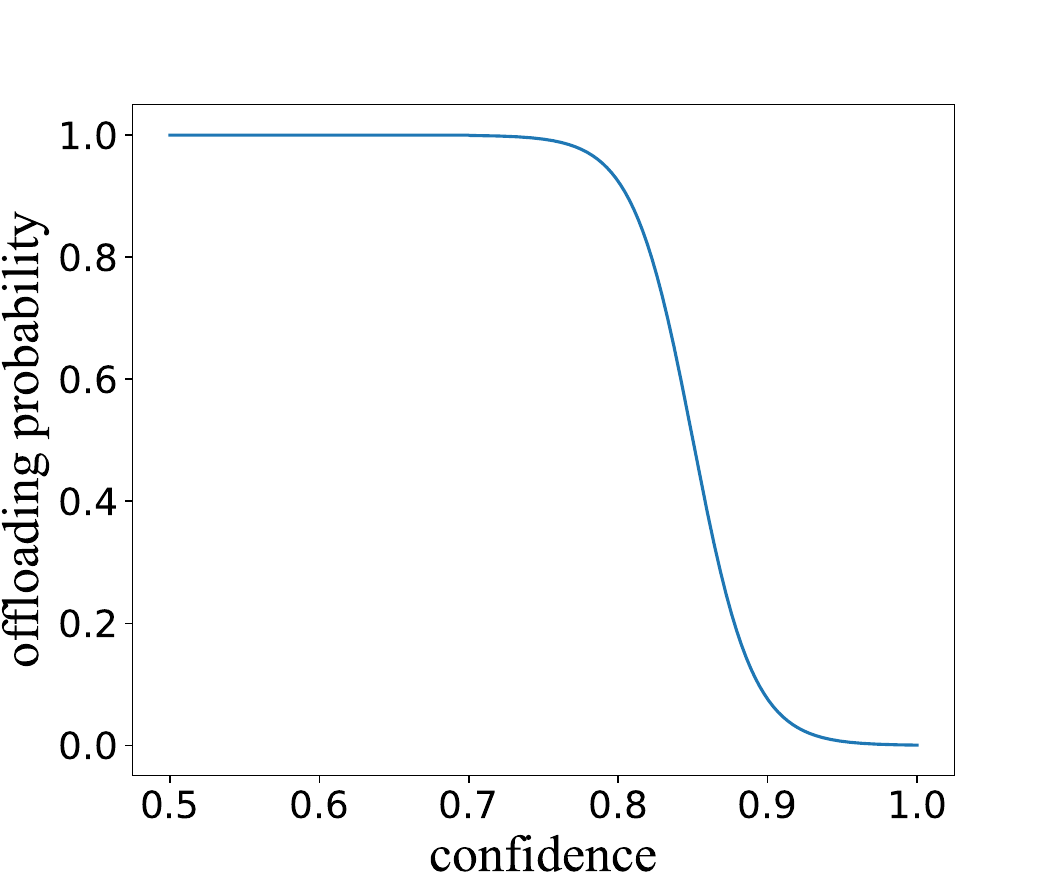}
    \caption{Confidence vs. offloading probability curve.}
    \label{fig:4}
\end{figure}

\autoref{fig:4} shows the relationship between the confidence and the probability of offloading in the task offloading decision, where the confidence threshold is $t$ = 0.7. It can be seen that the probability of offloading is 1 when the confidence level is lower than the threshold, while there is an S-shaped curve that adjusts the offloading decision with a smooth probability when the confidence level is higher than the threshold. Compared to a direct linear decrease in probability (i.e. the probability decreases in a straight line), this scheme allows more inference tasks to be returned earlier, thus reducing latency and achieving higher accuracy in the returned tasks. Specifically, for tasks with probabilistic offloading, lower confidence has a higher probability of offloading to improve accuracy, and higher confidence has a lower probability of offloading to improve overall efficiency\cite{wang2023tabi}.

\subsection{Attention-based Pruning}
Our system is designed to target models based on transformer architectures that provide discriminative services, including BERT and others. These models summarise the input, understand the meaning of the task text and make inferences. Such language models can perform a variety of NLP tasks based on encoded text representations, such as sentiment analysis, natural language inference, question answering, etc.

\begin{figure}[tp]
    \centering
    \includegraphics[width=5cm]{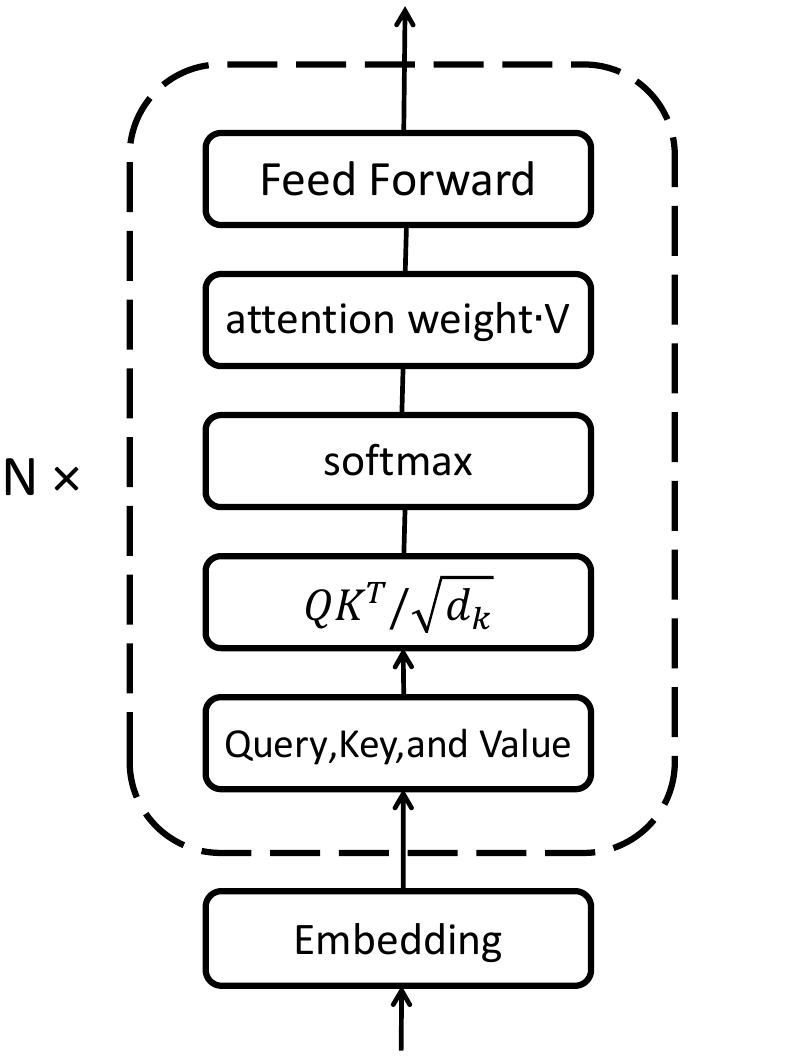}
    \caption{Simple structure of the transformer encoder.}
    \label{fig:5}
\end{figure}

The language model consists of a set of transformer encoder modules. In the input phase, words are digitised. Specifically, the language model segments the input sentence into sub-word tokens and maps these tokens to numerical vectors, which are then fed into the transformer modules in a process called tokenization. As shown in \autoref{fig:5}, The transformer encoder module is a special kind of neural network architecture. The input vector to each encoder module first passes through a self-attention module. The input to the self-attention mechanism consists of $Query$, $Key$, and $Value$ vectors, all of which have the same dimension. These input vectors are obtained by multiplying the original input vectors by their respective weight matrices. Within the self-attention module, the $Query$ vectors are dot-multiplied with the $Key$ vectors to compute attention scores. These scores are then normalized using the softmax function to produce attention weights. Subsequently, these weights are multiplied by the $Value$ vectors to generate the attention output. This process allows the model to reference information from all positions in the sequence when processing each input token, thereby capturing long-range dependencies within the sentence. The attention output is then passed through a Feed-Forward Network (FFN), which typically consists of two linear transformation layers and a non-linear activation function such as ReLU. This entire process is repeated across multiple Transformer encoder blocks, with each layer performing increasingly complex feature extraction and representation learning. As the input vectors traverse through the successive layers of self-attention and feed-forward networks, the model progressively extracts and integrates higher-level semantic information.

The language model with the transformer as its architecture is built on the basis of the attention mechanism. Attention encodes relationships between tokens, generates contextually relevant embeddings, and makes the model aware of relevant tokens when processing vectors. To calculate the attention scores, a scaled dot-product method, denoted as $QK^T / \sqrt{d_k}$, which evaluates the degree of relevance between two tokens. Subsequently, the softmax function is applied to these scores to accentuate the weights of highly correlated token pairs. The final step involves multiplying these attention weights with $V$, yielding the output of a single attention head. Specifically, it is calculated as follows:
\begin{equation}
Attention(Q,K,V)=softmax(\frac{QK^T}{\sqrt{d_k}})V
\end{equation}

The attention weight, given by $softmax(QK^T / \sqrt{d_k})$, transforms the ``attention" from one token to all other tokens within an input sequence into a normalized probability distribution that sums to 1. As a result, aggregating the attention weights from all tokens produces an importance vector, where tokens that garner more attention— and are thus deemed more significant— exhibit higher weights.

\begin{figure}[tp]
    \centering
    \includegraphics[width=5.5cm]{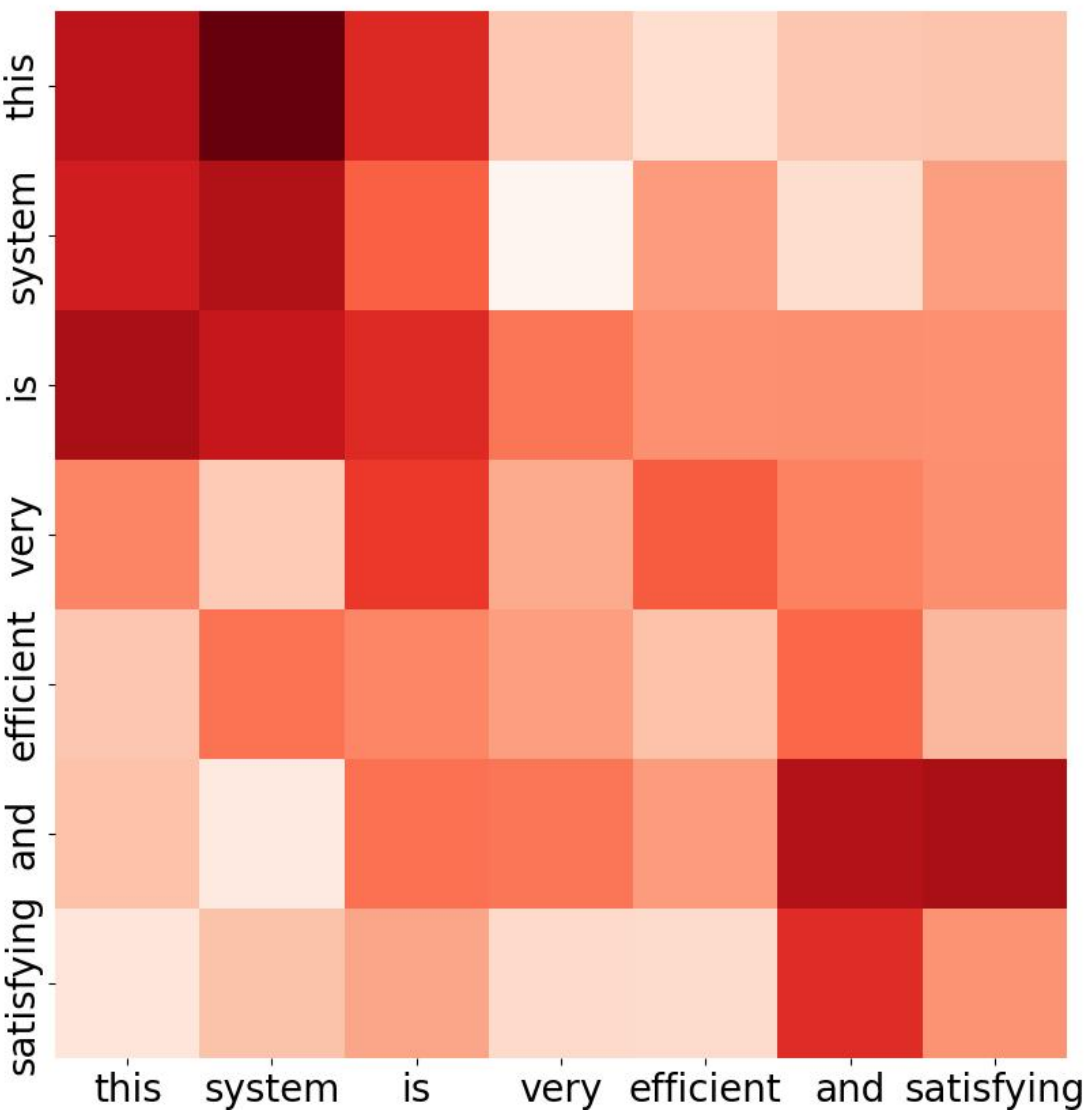}
    \caption{Examples of attention weights. Darker colours indicate more weight and importance.}
    \label{fig:6}
\end{figure}

As shown in \autoref{fig:6}, natural language contains many redundant words, such as prepositions and postpositions. While these words help convey grammatical relationships, they contribute minimally to natural language processing (NLP) tasks\cite{clark2019does},\cite{goyal2020power}. For example, words like ``in", ``at" and ``on" convey spatial or temporal information in sentences but are not significant for the core understanding and processing in certain NLP tasks.

By removing these redundant words, we can significantly accelerate inference. This is because the computational complexity of the attention mechanism increases quadratically with the sequence length\cite{teerapittayanon2016branchynet}. In other words, the longer the sequence, the greater the computational load. Therefore, by reducing the sequence length through the elimination of unimportant words, we can decrease computational complexity and improve processing speed. This approach not only accelerates the inference process but also enhances the overall efficiency and performance of the system to some extent.

Language models rely on the attention mechanism to determine the semantic importance of each sub-word token. The attention mechanism calculates attention weights, indicating the relative importance of each sub-word in understanding the sentence's semantics. To assess the importance of a query, we extract and accumulate the attention weights from each layer of the model. These attention weights are intermediate variables generated during inference, making the accumulation process both fast and efficient.

Our pruning process can be understood as follows: if the importance of the token $y$ to be pruned is greater than the pruning threshold (the pruning threshold is obtained by multiplying the pruning coefficient $\alpha$ by the average importance of the sentence $ave$), the value of the equation is 1, which means that the token is retained, whereas if the importance of the token $y$ to be pruned is less than or equal to the pruning threshold, the value of the equation is 0, which means that the token is deleted. The process is formally expressed by the following equation:
\begin{equation}
pruning(y)=\left\{
	\begin{aligned}
	1 \quad imp_y > \alpha \cdot ave\\
	0 \quad imp_y \leq \alpha \cdot ave\\
	\end{aligned}
	\right
	.
\end{equation}
this pruning threshold is based on the average importance of the sentence and can be adjusted to accommodate different inference environments.

When using language models, having only the token pruning mask is insufficient because different models often employ various tokenizers to handle and represent tokens. Tokenizers segment text into smaller units and convert them into numeric representations. Additionally, different tokenizers might map the same word to different numeric IDs. Specifically, for some words with prefixes and suffixes, certain models treat them as a single token, while other models separate the prefixes and suffixes from the word root, treating them as multiple tokens. Therefore, when transferring data between different models, the only universal data passing interface is the raw text itself.

Our system does not specify the method of token segmentation for models at the design stage. Therefore, to ensure that our attention-based pruning accommodates the characteristics of different models, we establish the following rules for implementation:

\begin{itemize}
\item[1)]
If the token segmentation of the current and subsequent models is the same, pruning is performed according to the original pruning method.
\item[2)]
If the previous model does not differentiate between prefixes and suffixes but the subsequent model does, the word retained (or pruned) by the previous model will be entirely retained (or pruned), including the prefix, suffix, and root.
\item[3)]
If the previous model differentiates between prefixes and suffixes but the subsequent model does not, the word will only be entirely pruned if both the prefix, suffix, and root are pruned; otherwise, the entire word will be retained.
\end{itemize}

\subsection{Ensemble}
For the offloaded inference tasks, we do not rely on the final model's output as the ultimate result. Instead, we adopt a weighted aggregation method, where the inference results from each model are combined based on their respective weights. This approach introduces no additional computational overhead or time costs, as obtaining the results and performing the aggregation is both simple and efficient.


According to the above design, the final result of a task $x$ that has undergone inference through $m$ models is calculated as follows:

\begin{equation}
result(x)=\sum_{i=1}^m w_i \cdot softmax_i(x)
\end{equation}
where:
\begin{equation}
\sum_{i=1}^m w_i = 1
\end{equation}
where $w_i$ is the weight of the $i$-th model, and $softmax_i$ is the inference result of the $i$-th model. Obviously, the sum of all weights should be 1. This method of computing results can partially offset the accuracy loss caused by pruning and other factors. The weight of each model in the aggregation process is determined by its intrinsic attributes, such as accuracy. For instance, in an inference task, higher-level models are assigned greater weight, thereby exerting a larger influence on the final result.

\subsection{Optimized Early Exit}
In our system, we have implemented an early exit technique to optimize the inference speed of individual models. Upon satisfying the exit conditions, the output from the current layer is immediately utilized as the final result. This approach obviates the need for further computation in the remaining layers, thereby significantly reducing the wastage of computational resources.

We refer to previous research on early exit methods for DNN-based models. Specifically, the exit problem is divided into two parts: 1) When to exit and 2) How to exit.

\begin{figure}[tp]
    \centering
    \includegraphics[width=8.5cm]{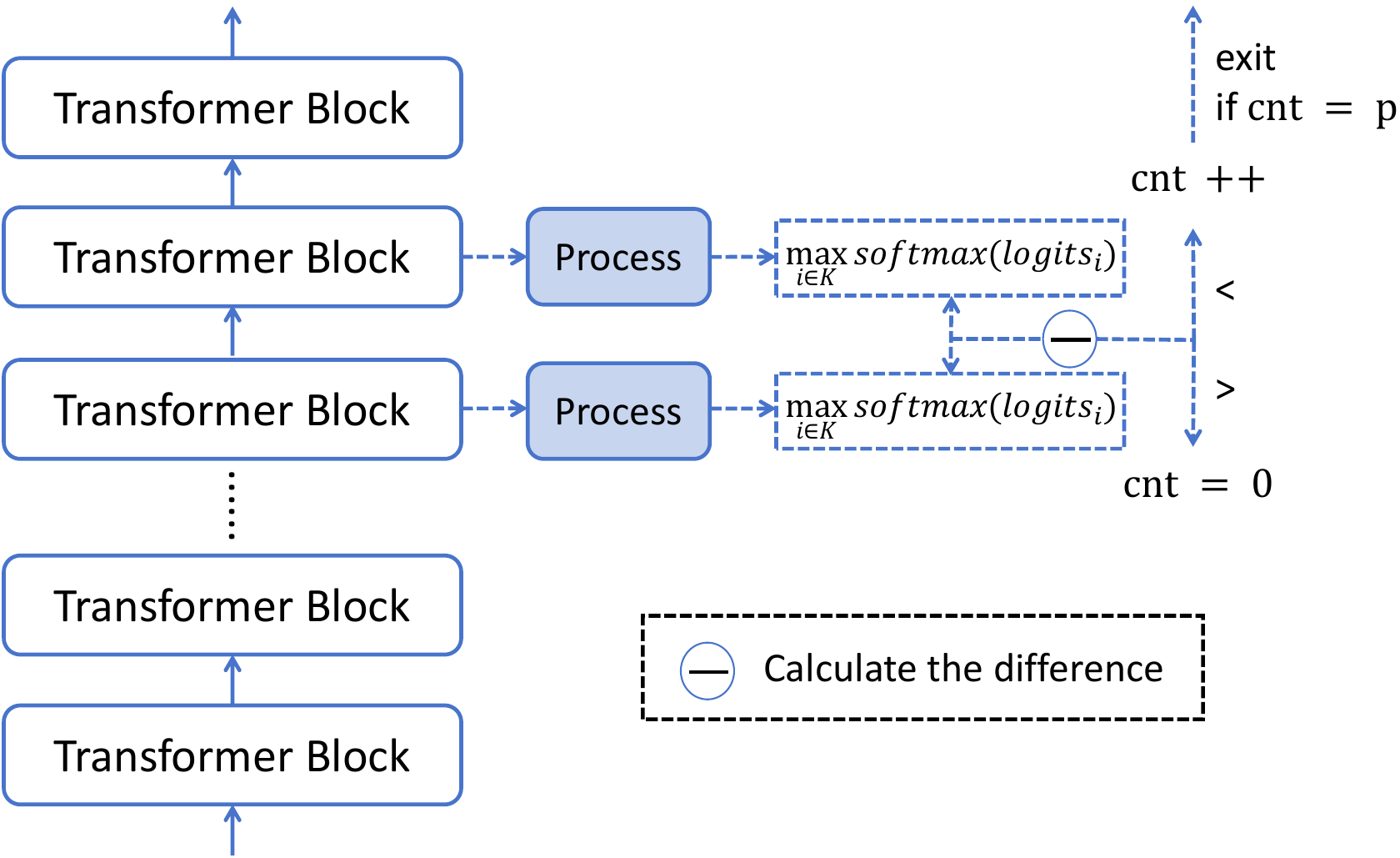}
    \caption{Optimized early exit.}
    \label{fig:7}
\end{figure}

\autoref{fig:7} shows the design of the early exit mechanism for individual model acceleration in our system. The $Process$ module in the figure has a similar structure to the classification layer and is able to transform the intermediate layer values into intermediate classification results. The system compares the similarity of two adjacent layers, formally expressed as follows:
\begin{equation}
diff(x)=\lvert  \max softmax_k - \max softmax_{k-1} \rvert
\end{equation}
where $softmax_k$ and $softmax_{k-1}$ are the results for layer $k$ and layer $k-1$ respectively. 

After obtaining the difference values of the two adjacent layers, we compare them with a threshold $\tau$ and introduce a counter $cnt$. If the difference is large, we set the counter $cnt$ to zero. If the difference is small, representing a possible redundancy of operations in the subsequent layers, we add 1 to the counter:
\begin{equation}
cnt=\left\{
	\begin{aligned}
	cnt + 1 \quad diff(x) < \tau\\
	0 \quad diff(x) > \tau\\
	\end{aligned}
	\right
	.
\end{equation}

Until the counters accumulate to a value equal to the exit threshold $p$, the system makes an exit decision and takes the result of the last layer before exit as the inference result for that model.

If the above exit condition cannot be met, the model passes through all the transformer blocks. This design allows individual models to achieve a better trade-off between inference time and accuracy, as well as improved inference efficiency for the overall inference system.

\section{Evaluation}

In this section, we conduct a thorough evaluation of our system's performance through a series of experiments. The evaluation focuses on two key metrics: inference accuracy and inference time, under varying parameter settings, such as different early exit and confidence thresholds. Multiple pre-trained models are tested in a high-performance computing environment to assess their performance.


\subsection{Experimental Setup}

To evaluate the performance of our proposed collaborative inference system, we conduct experiments under various configurations. Our experiments mainly focus on inference accuracy and inference time.

\subsubsection{Dataset}

We primarily use the IMDB dataset, a classic sentiment analysis dataset containing movie review texts and their corresponding sentiment labels. This kind of data set can make the results more general.

\subsubsection{Models}

We select several popular BERT models to evaluate our system, including:
\begin{itemize}
    \item \texttt{bert-base-uncased}: it belongs to the BERT series, has a base size, and features uncased preprocessing.
    \item \texttt{bert-large-uncased}: it belongs to the BERT series, has a large size, and features uncased preprocessing.
    \item \texttt{bertweet}: it belongs to the BERTweet series, has a base size, and is specifically trained on Twitter data.
\end{itemize}

Each model is pre-trained and fine-tuned on the IMDB dataset for the sentiment analysis task.

\subsubsection{Environment}

Experiments are conducted in a high-performance computing environment with the following configuration: an Ubuntu 22.04 operating system, Python 3.12, PyTorch 2.3.0, CUDA 12.1, a single RTX 4090 GPU with 24GB of memory, a 16-core Intel(R) Xeon(R) Gold 6430 CPU, and 120GB of RAM. Additionally, we incorporate communication delays from a real network, accounting for factors such as network congestion, signal interference, and varying bandwidth. This enables a more accurate simulation of the communication challenges encountered in mobile network environments, where latency can significantly affect the performance of distributed inference systems.

\subsubsection{Parameters}

The key parameters of the system are configured as follows: the Early Exit Threshold ($\tau$) is set to 0, 0.01, 0.0001, and 0.00001, while the Confidence Threshold is set to 0.7, 0.8, and 0.9.

\subsection{Inference Accuracy and Time}

We test the system's inference accuracy and time under different configurations. The results are shown in \autoref{table:results}. Setting the Early Exit Parameter to 0 indicates that early exit is disabled, thereby reproducing the results achieved by Tabi\cite{wang2023tabi}.

\begin{table}[tp]
    \centering
    \caption{Inference accuracy and time.}
    \label{table:results}
    \resizebox{\columnwidth}{!}{
        \begin{tabular}{|c|c|c|c|c|}
            \hline
            \textbf{Early Exit Parameter ($\tau$)}& \textbf{Confidence Threshold} & \textbf{Time (ms)} & \textbf{Accuracy} \\
            \hline
            0 & 0.9 & 534.7 & 0.9685 \\
            0.01 & 0.9 & 412.21 & 0.7856 \\
            0.0001 & 0.9 & 443.39 & 0.8236 \\
            0.00001 & 0.9 & 502.98 & 0.9195 \\
            \hline
            0 & 0.8 & 523.93 & 0.9334 \\
            0.01 & 0.8 & 401.82 & 0.7664 \\
            0.0001 & 0.8 & 439.92 & 0.8121 \\
            0.00001 & 0.8 & 497.71 & 0.9012 \\
            \hline
            0 & 0.7 & 510.81 & 0.9226 \\
            0.01 & 0.7 & 398.53 & 0.7162 \\
            0.0001 & 0.7 & 425.56 & 0.7833 \\
            0.00001 & 0.7 & 473.28 & 0.8610 \\
            \hline
        \end{tabular}
    }
\end{table}

\subsubsection{Accuracy Analysis}

When the confidence threshold is set to 0.9, accuracy shows varying trends as the early exit threshold ($\tau$) changes. As $\tau$ decreases, accuracy initially drops before increasing, indicating that the early exit strategy can help maintain relatively high accuracy under a high confidence threshold.

At lower confidence thresholds, the overall accuracy is lower than at 0.9. Specifically, accuracy is lowest when the confidence threshold is 0.7, highlighting that a lower probability threshold significantly impacts inference accuracy.


For a confidence threshold of 0.9 and an early exit parameter of 0, accuracy is 0.9685. However, when the early exit parameter increases to 0.01, accuracy drops to 0.7856, demonstrating the model's sensitivity to early exit. At lower thresholds such as 0.7 and 0.8, similar trends are observed, though the baseline accuracy is lower.


\subsubsection{Time Analysis}
As the early exit threshold increases (i.e., as $\tau$ decreases), the inference time becomes shorter. This is because the early exit strategy allows the inference process to terminate sooner, reducing overall computation time. While the confidence threshold has a less pronounced effect on inference time, the general trend shows that lower confidence thresholds lead to shorter inference times.


For an early exit parameter of 0 and a confidence threshold of 0.9, the inference time is 534.7 ms. When the early exit parameter decreases to 0.01, the inference time drops to 412.21 ms, demonstrating the efficiency gains from the early exit mechanism. This trend is consistent across other confidence thresholds, further highlighting the role of early exit in accelerating the inference process.


\subsubsection{Overall Performance}

\textbf{Best Combination}: Balancing both accuracy and inference time, the optimal combination is a confidence threshold of 0.8 and an early exit threshold of 0.00001. This configuration maintains an accuracy rate of no less than 80\% while reducing inference time by approximately 17\%. \textbf{Balance}: If accuracy is the priority, we recommend using a confidence threshold of 0.9 with an early exit threshold of 0.00001. On the other hand, if inference time is more critical, a confidence threshold of 0.7 with an early exit threshold of 0.01 should be used.

With a confidence threshold of 0.8 and an early exit parameter of 0.00001, the system achieves an accuracy of 0.9012 and an inference time of 497.71 ms, representing a good balance between accuracy and efficiency. When accuracy is prioritized, setting the confidence threshold to 0.9 and the early exit parameter to 0.00001 results in an accuracy of 0.9195 but a longer inference time of 502.98 ms. Conversely, when prioritizing inference time, a confidence threshold of 0.7 and an early exit parameter of 0.01 yields the shortest time of 398.53 ms, though at the cost of lower accuracy (0.7162).


\subsection{Detailed Analysis of Experimental Results}

\subsubsection{Comparison Analysis}

\begin{figure}[tp]
    \centering
    \includegraphics[width=7cm]{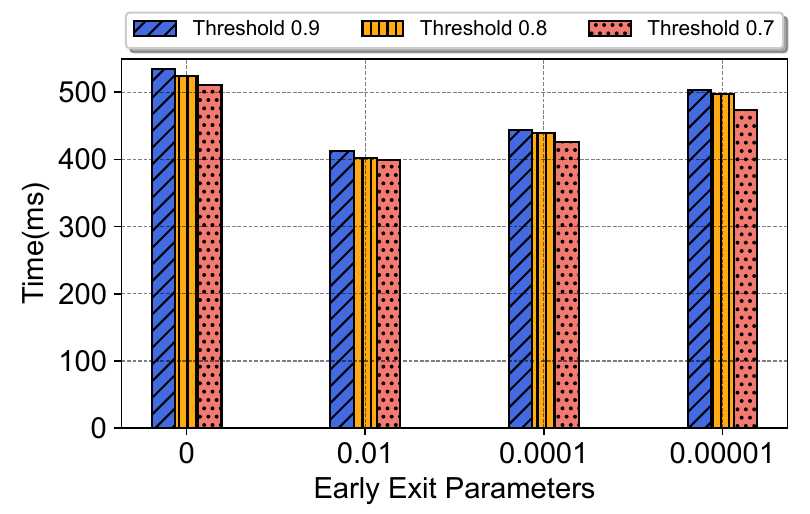}
    \caption{Comparison of time.}
    \label{fig:9}
\end{figure}

As shown in \autoref{fig:9}, inference time is significantly influenced by the early exit threshold across all probability thresholds. This pattern highlights the robustness of the early exit mechanism in optimizing computational efficiency without a substantial loss in accuracy.

\begin{figure}[tp]
    \centering
    \includegraphics[width=7cm]{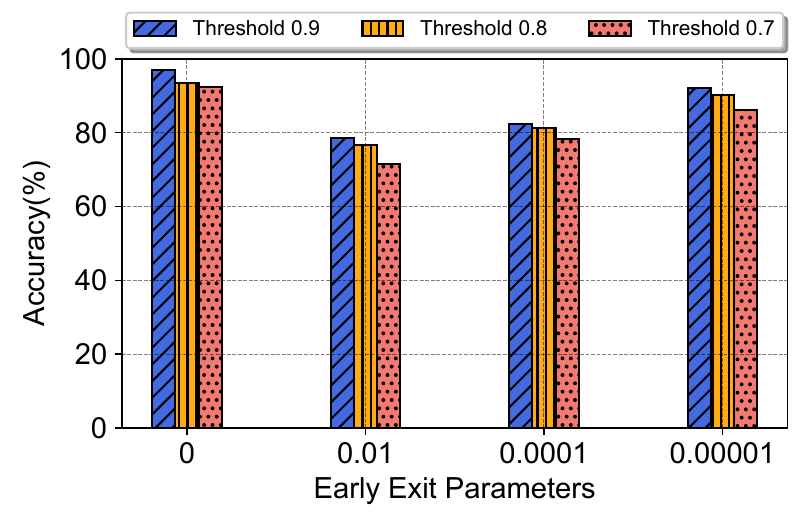}
    \caption{Comparison of accuracy.}
    \label{fig:8}
\end{figure}

From \autoref{fig:8}, it is evident that the confidence threshold plays a crucial role in determining the baseline accuracy. A higher threshold results in higher accuracy but may increase inference time, while a lower threshold can reduce the inference time at the expense of accuracy.


Comparing rows with the same early exit parameter but different confidence thresholds reveals that higher thresholds, such as 0.9, consistently yield better accuracy, albeit with longer inference times. This underscores the importance of fine-tuning both parameters to strike the right balance between accuracy and efficiency.


\subsubsection{Implications for Real-World Applications}

The experimental results demonstrate that our proposed collaborative inference system can effectively balance accuracy and inference time, making it adaptable to a wide range of application scenarios.  For applications requiring high accuracy, such as critical decision-making systems, we recommend setting a high confidence threshold and a low early exit threshold.  Conversely, for applications needing rapid responses, such as real-time monitoring systems, a lower confidence threshold and a higher early exit threshold may be more suitable. Additionally, our system exhibits strong scalability and versatility, as it is not tied to any specific model, task, or network configuration.  The system framework is designed to seamlessly operate across diverse environments, including varying network conditions, device capabilities, and model architectures.  This flexibility allows it to accommodate different inference requirements.  By fine-tuning key system hyperparameters, such as the confidence and early exit thresholds, the framework can be customized to optimize performance according to specific application demands, ensuring efficiency and effectiveness across a variety of use cases.


\section{CONCLUSION}
In this paper, we propose a collaborative inference system designed for next-generation networks and communication, addressing the challenges posed by uneven computational resources in real-world mobile environments. Our system facilitates multi-level inference and computation offloading, enabling a dynamic balance between inference accuracy and delay. Through extensive experimentation, we demonstrate that our system can reduce inference latency by up to 17\%, while maintaining the accuracy of the predictions. This performance improvement is achieved without compromising the quality of the inference process, making our approach more efficient than existing solutions. Future work will consider more dynamic systems to adapt to more variable network environments.


\bibliographystyle{IEEEtran}
\bibliography{IEEEabrv,mylib}

\vfill

\end{document}